\documentclass[namedreferences]{kluwer}    
\usepackage{graphicx}

\newdisplay{guess}{Conjecture}

\begin{document}                                       
                                            
\begin{article}
\begin{opening}         
\title{Confidence Limits of Evolutionary Synthesis Models\thanks{Work
            motivated by the Granada Euroconference. We want to acknowledge the LOC for financial support.}}
\author{Miguel \surname{Cervi\~no}}  
\institute{Max--Planck--Institut f\"ur Extraterrestrische Physik}
\author{Valentina \surname{Luridiana}}  
\institute{Instituto de Astrof\'\i sica de Andaluc\'\i a (CSIC)}

\runningauthor{M. Cervi\~no \& V. Luridiana}
\runningtitle{Confidence Limits of Synthesis Models}
\date{October 30, 2001}

\begin{abstract}
The probabilistic nature of the IMF in stellar systems implies that clusters
of the same mass and age do not present the same unique values of their
observed parameters. Instead they follow a distribution. We address the
study of such distributions in terms of their confidence limits that can be
obtained by evolutionary synthesis models. These confidence limits can be
understood as the inherent uncertainties of synthesis models. We will
compare such confidence limits arising from the discreteness of the number 
of stars obtained with Monte Carlo simulations with the dispersion resulting
from an analytical formalism. We give some examples of the effects on the
kinetic energy, V--K, EW(H$\beta$) and multiwavelength continuum.

\end{abstract}

\end{opening}           

\section{Introduction}

In recent years, several efforts have been dedicated to improve our
understanding of stellar evolution with more detailed and complete
theories; at the same time, more powerful observatories have been developed
to test the theory.  However, an intermediate tool is necessary to link
these pieces of information when we deal with systems in which only the
integrated light of stellar populations (and their by-products, like the
emission line spectrum) is available: this tool are synthesis models.
Recently, it has been established how the input libraries affect the
predictions of synthesis models (see \opencite{Bru01a} or 
\opencite{Car00} as examples).
From the theoretical point of view, there are still several open questions
in the modelization of stellar clusters by evolutionary synthesis
codes. One of the most important ones is related with the conservation of
energy and the Fuel Consumption Theorem established by
\inlinecite{RB86} (see also \opencite{MG01} for the link of chemical with
spectrophotometric models). Other questions, related with ``technical''
details in the isochrones computation can be found in
\inlinecite{Cetal01a}. We also refer to the contribution of S. Yi in these
proceedings.

In addition to those listed, there is still a source of uncertainty
arising from the use of the Initial Mass Function (IMF) and the effect
of the discreteness in the number of stars in the models results
(\opencite{Buz89}; \opencite{SF97}; \opencite{LM99}; \opencite{CLC00};
\opencite{Plu01}; \opencite{Bru01b}; \opencite{Cetal01b}).  In this
paper we present our current understanding of the dispersion
introduced in the results of evolutionary synthesis models by the
discreteness of the stellar population for a given IMF.

\section{The modelization of real star forming regions}

Using a careful analysis of the stars in the solar neighborhood,
\inlinecite{Kro01} shows that the Salpeter IMF is compatible with
observations if stochastic effects are taken into account.  Taking into
account the discreteness of the stellar population, the predictions of any
model that relies on an IMF are only exact {\it under the assumption of an
infinite number of stars}. Otherwise, they only give a {\it mean value} of
a probability distribution. The relevance of such fluctuations in the
results of synthesis models is obvious in the case of massive stars and
young clusters, {\it but they also affect the models of older clusters
dominated by the emission of low-mass stars} since small variations in the
initial mass/number of stars in a given mass range, can produce different
numbers of, e.g., AGB stars at a given age, which in turn produce large
variations in the resulting colors (see \opencite{SF97} and
\opencite{Bru01b} as examples).

So, for the comparison of models with observational data it is necessary to
obtain not only the mean value of the observables, but also, at least, the
corresponding dispersion of the computed observables due to the
discreteness of the stellar population.  Such dispersion can be evaluated
theoretically in function of the effective number of stars,
$N_\mathrm{eff}$ (\opencite{Buz89}; see also \opencite{Cetal01b}):

\begin{equation}
\frac{\sigma_{L}}{<L>} = \frac{1}{\sqrt{N_\mathrm{eff}(L)}},
\label{eq:neff}
\end{equation}

\begin{figure}
\centerline{\includegraphics[width=30pc]{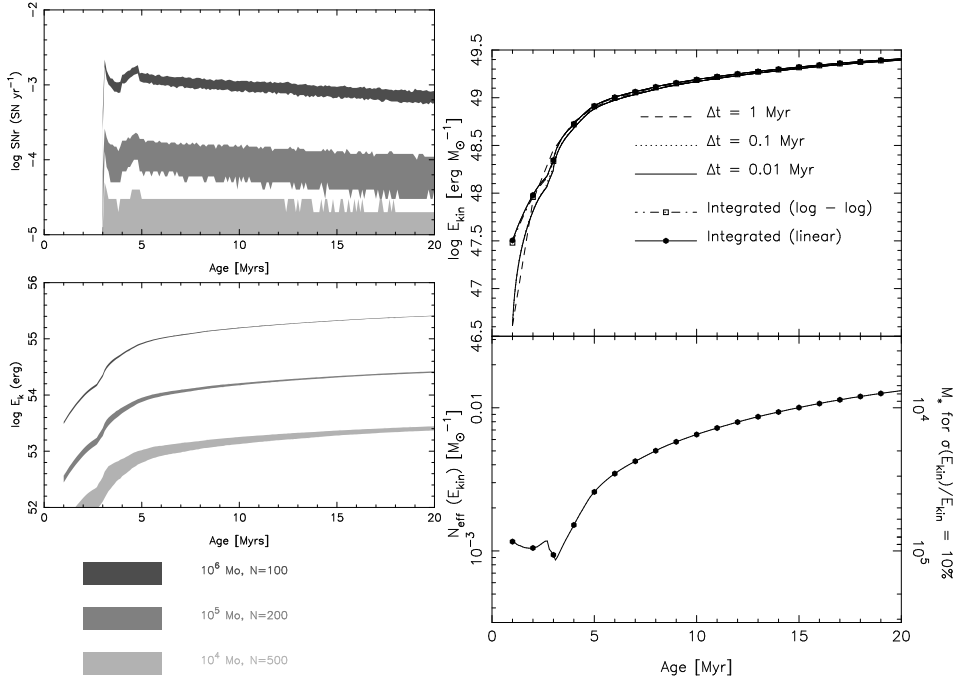}}
\caption{{\it Left:} Monte Carlo 90\% Confidence Levels for the SN rate and
the Kinetic for different amount of stars. {\it Right:} Analytical
simulations and the corresponding $N_{\mathrm{eff}}(E_\mathrm{K})$. Note
than $N_{\mathrm{eff}}(SNrate)$= SNr by definition. }
\label{fig:ek}
\end{figure}

\begin{figure}
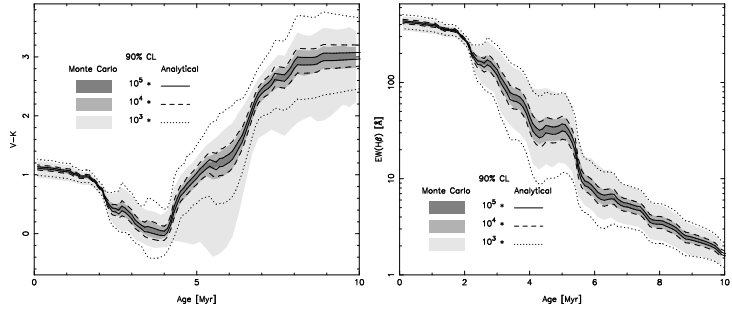

\centerline{\includegraphics[width=9.5pc,angle=270]{./cervino.m.fig2.eps}
\includegraphics[width=9.5pc,angle=270]{./cervino.m.fig3.eps}}
\caption{Monte Carlo and analytical 90\% Confidence Levels for V--K and the
EW(H$\beta$) in emission.}
\label{fig:VKHb}
\end{figure}

\begin{figure}
\centerline{\includegraphics[width=12pc,angle=270]{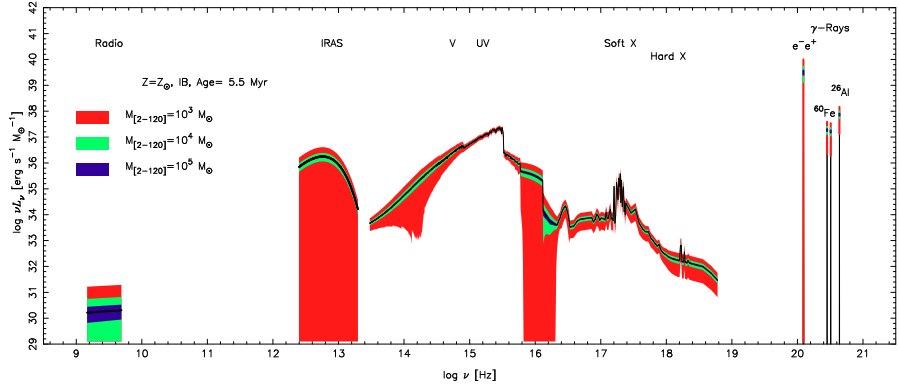}}
\caption{Analytical 90\% Confidence Level for the multiwavelength spectrum
for a 5.5 Myr old burst.}
\label{fig:multi}
\end{figure}

As \inlinecite{Buz89} highlighted, $N_\mathrm{eff}$ is not a real number of
stars, but rather a rough estimate of the number of stars contributing to a
given variable.  In Figures \ref{fig:ek} to \ref{fig:multi} we show several
examples of the 90\% confidence level of a large number of Monte Carlo
simulation and their comparison with the relative dispersion obtained by
the analytical formalism. All models assume an instantaneous burst of star
formation and solar metallicity evolutionary tracks. Other examples can be
found in \citeauthor{CLC00} (\citeyear{CLC00}, \citeyear{Cetal01a},
\citeyear{Cetal01b}).

\section{Conclusions}

We have shown that the effects of fluctuations in the number of
stars due to the stochastic nature of the stellar formation process and the
discreteness of the stellar populations produce a dispersion in the
predictions of evolutionary synthesis models, and that such dispersion may
be much larger than the observational errors. The dispersion can be
evaluated theoretically, and it can be used as an observable.  The
application of this ideas may help to improve the understanding of other
astrophysical problems, for example: (i) Is it really necessary a IMF slope
different from Salpeter's?  (ii) Is it possible to explain the observed
dispersion of chemical abundances by including the IMF fluctuations in
chemical evolution models?.  (iii) How much the underlying probability
distribution of luminosities affects the corresponding colors?...




\end{article}
\end{document}